\documentclass[aps, prb, twocolumn, groupedaddress, showkeys]{revtex4}
\usepackage[dvips]{color}
\usepackage{xcolor}
\usepackage{amsfonts}
\usepackage{amsmath}
\usepackage{amssymb}
\usepackage{graphicx}
\usepackage{mathrsfs}
\usepackage{bbm}
\usepackage{bm}
\usepackage{multirow}
\usepackage[sort&compress]{natbib}
\usepackage{dcolumn}%

\begin{document}

\newcommand{\dd}{d}
\newcommand{\pd}{\partial}
\newcommand{\myU}{\mathcal{U}}
\newcommand{\myr}{q}
\newcommand{\Urho}{U_{\rho}}
\newcommand{\myalpha}{\alpha_*}
\newcommand{\bd}[1]{\mathbf{#1}}
\newcommand{\Eq}[1]{Eq.~(\ref{#1})}
\newcommand{\Eqn}[1]{Eq.~(\ref{#1})}
\newcommand{\Eqns}[1]{Eqns.~(\ref{#1})}
\newcommand{\Figref}[1]{Fig.~\ref{#1}}
\newtheorem{theorem}{Theorem}
\newcommand{\me}{\textrm{m}_{\textrm{e}}}
\newcommand{\sgn}{\textrm{sign}}
\newcommand*{\bfrac}[2]{\genfrac{\lbrace}{\rbrace}{0pt}{}{#1}{#2}}

\title{Minimal cosmological masses for photons and gluons}

\author{Lorenzo Gallerani Resca}
\email{resca@cua.edu}
\homepage{http://physics.cua.edu/people/faculty/homepage.cfm}

\affiliation{Department of Physics and Vitreous State Laboratory, 
The Catholic University of America,
Washington, DC 20064}

\date{\today}

\begin{abstract}

I conjecture non-zero photon and gluon bare masses derived from a cosmological scale. That entails the existence of photon and gluon Bose-Einstein (BE) condensates in a comoving Friedmann-Lemaitre-Robertson-Walker geometry. I derive tantalizing results that set the inter-particle distance in these BE condensates almost at the nucleon scale and a corresponding critical temperature of condensation almost at the end of the quark epoch or the beginning of quark-gluon confinement. My estimates for particle mass and number density in these BE condensates suggest remarkable relations among fundamental constants, $h, c, G, \Lambda$, at the most microscopic and cosmological scales of quantum and relativity theories.  

\end{abstract}





\maketitle

The theory of quantum mechanics (QM) and quantum fields (QFT) on the one hand, and the theory of general relativity (GR) on the other hand, form the two greatest pillars of modern physical science. Each theory accounts for phenomena on a vast range of scales, where each theory has been confirmed with astonishing precision by experiments and observations conducted with instrumentation of unprecedented sophistication or size. Yet ranges of primary success of each theory do not quite overlap. A unified theory that may encompass `all' scales of the observable universe remains to be conclusively established. That poses more and more formidable challenges, but scientific progress advances relentlessly.

In this paper I examine a relation between the smallest of scales for particle masses and the largest of scales for the size of the observable universe. That leads to the hypothesis that `dark energy' associated with the cosmological constant, $\Lambda$, consists of Bose-Einstein (BE) condensates of massive photons or gluons at rest in the freely falling comoving frame of the Friedmann-Lemaitre-Robertson-Walker (FLRW) geometry.\cite{Hobson, Schutz2Ed} 

Namely, 

\begin{equation}\label{fundamental}
\lambda_g = \frac{h}{m_g c} \sim  \sqrt{\frac{1}{\Lambda}} \simeq 10^{10} \mathrm{ly}  .
\end{equation}  
Here, $\lambda_g$ refers to the Compton wavelength of either massive photons or gluons. While such a wavelength for electrons, $\lambda_e$, was originally introduced by Compton to characterize their scattering of massless photons, $\lambda_m = \frac{h}{m c}$ maintains a more general and profound meaning in QFT. Indeed, $\lambda_m$ enters the uncertainty principle when creation of a rest mass with energy $m c^2$ is required by special relativity (SR). Correspondingly, $\lambda_m$ enters explicitly or implicitly most fundamental QM and QFT equations for massive particles and fields. Thus, \Eq{fundamental} provides a connection between a most fundamental microscopic uncertainty relation and a most fundamental cosmological term in the observable universe.

In \Eq{fundamental}, the `comparable' $\sim$ sign indicates that numerical coefficients, most likely different for photons and gluons, should enter into more precise relations if or when more quantitative theories are developed. At this stage, connecting most fundamental constants of quantum mechanics and relativity, such as $h$, $c$ and $\Lambda$, across most microscopic and cosmological scales, provides just a tantalizing clue, somewhat complementary to that of Planck scale, but in the opposite infra-red limit. 

Assuming \Eq{fundamental}, the conceivably minimal cosmological rest-energy mass of gauge bosons is of the order of
\begin{equation}\label{mass}
m_g c^2 = \frac{h c}{\lambda_g} \sim  h c \sqrt{\Lambda} \simeq 1.3 \mathrm{x} 10^{-32} \mathrm{eV} \simeq 1.4 \mathrm{x} 10^{-41} m_p c^2 ,
\end{equation}  
where $m_p c^2 \simeq 938 \mathrm{MeV}$ represents the proton rest-energy mass.

Upper bounds to the photon rest-energy mass, $m_{\gamma} c^2$, have been determined by a variety of experiments, observations or inferences. Perhaps most stringent upper limits of astrophysical origin require $m_{\gamma} c^2 < 3 \mathrm{x} 10^{-27} \mathrm{eV} $.\cite{Quigg, Jackson, Dolgov, Pani} Whatever the case, any such upper bound is several orders of magnitude greater than the conceivable minimum of \Eq{mass}.

A heuristic consideration of \Eq{fundamental} derives from the fact that mass is a continuous variable with physical dimensions and without quantization. Masses are neither large nor small in the absolute. Mass scaling poses formidable conceptual and technical challenges to renormalization in QFT. A `zero mass' has no scale and it is not even a mass: it is a `zero,' as in the ideals of mathematics. An `infinitesimally small' or a `vanishing' mass is an entirely different concept. In GR, `test particles' with vanishing mass all follow the same time-like geodesics. Null geodesics are altogether different, although needed to represent a limiting condition. There is however no definitive requirement in GR that massless particles should exist in reality. In fact, Einstein disliked the very concept of `particle' and he finally succeeded with others in proving that geodesic equations ultimately derive from GR field equations themselves.\cite{Rindler} So, in either QFT or GR it would be more consistent to assume that all `particles' have some finite mass, or none would, I suppose.

A powerful principle of gauge invariance in QFT initially assumes that particles are massless. However, through ingenious mechanisms of spontaneous symmetry breaking, massive particles are generated, except for photons and gluons, which are maintained as massless gauge bosons in the standard model (SM) of elementary particles.\cite{Quigg, Mandl} The SM is a theory developed by many masters over several decades, making remarkable predictions and receiving extraordinary experimental confirmations, including the most recent discovery of a likely Higgs boson, which is indeed responsible for a spontaneous symmetry breaking mechanism of mass generation. Still, the SM assumes that space and time are \textit{infinite}, without regard for GR. However, the observable universe appears to be finite, while expanding with an acceleration consistent with a GR cosmological term or a `dark energy,' according to the standard model of cosmology or $\Lambda$-CDM.\cite{Hobson, Schutz2Ed} Thus, spontaneous symmetry breaking and other theorems in QFT that require the existence of massless gauge bosons in the SM cannot definitively rule out a relation such as \Eq{mass}.

Exact symmetries requiring massless photons and gluons in QFT are needed to provide exact conservation of charge and color. Still, minuscule violations of such conservation laws cannot be ruled out in a more general QFT applicable to a cosmological scale, if or when that theory is developed. Exactly massless photons and gluons are also required to make gauge QFT's renormalizable. On the other hand, renormalization theory has already been developed to such a degree of complexity that it is hard to say whether or not that could be further generalized to accommodate minuscule mass terms violating gauge invariance on a cosmological scale. In fact, such terms may even help to cut off infra-red divergencies.\cite{Quigg, Mandl} 

In Einstein's formulation of SR, constancy of the `speed of light' or `celerity,' $c$, is a central postulate assumed to yield and hold for all Lorentz transformations of the infinite Minkowski space-time. In particular, constancy of the `speed of light' is needed in SR to synchronize all clocks at rest in any given Lorentzian frame. Yet, SR cannot account for gravity and cannot be globally exact, unless there is no matter nor energy up to infinity. When any of that is introduced anywhere, constancy of the speed of light is valid only as a first-order approximation locally in a freely falling Lorentzian frame. Therein, cosmological corrections to $c = \mathrm{const}$ for massive photons would be of the order of $\frac{1}{2} {\Lambda \lambda^2} \simeq 10^{-39}$ for wavelengths $\lambda$ of visible light, say. Thus, GR also may not definitively rule out a relation such as \Eq{mass}.

Presently, the `dark energy' density of our accelerating observable universe is estimated to be\cite{Hobson, Schutz2Ed} 

\begin{equation}\label{darkenergy}
\rho_{\Lambda} = \frac{c^{4} \Lambda}{8 \pi G} \simeq 3.6 m_p c^2 / m^3 .
\end{equation}  
According to \Eq{fundamental}, there may then be a BE condensate in the FLRW geometry with a number density
\begin{equation}\label{numberdensity}
n_g = \frac{\rho_{\Lambda}} {m_g c^2} \sim \frac{c^3 \sqrt{\Lambda}} {8 \pi h G} = \frac{\sqrt{\Lambda}} {8 \pi l_P^2} \simeq 2.6 \mathrm{x} 10^{41} / m^3 ,
\end{equation}  
where $l_P = \sqrt{ \frac{h G} {c^{3}} } \simeq 4 \mathrm{x} 10^{-35} m$ denotes Planck length.

According to that estimate, the average distance between massive bosons in a BE condensate phase is of the order of
\begin{equation}\label{averagedistance}
d_g \sim  {n_g^{-1/3}}  \sim 1.6 \mathrm{x} 10^{-14} m = 16 fm .
\end{equation}  

Since the size of nucleons is of the order of $1fm$, we expect that quark-gluon confinement begins at that scale, while asymptotic freedom may begin at the scale of $0.1fm$. Given cancellations in the tens of orders of magnitude in performing my estimates, it is not so disappointing to find that $d_g$ is only about 16 times greater than one fermi. Perhaps BE massive gluons, randomly distributed at rest on the FLRW geometry, attract isotropically one another, thus contributing to the negative pressure $p_\Lambda = - \rho_{\Lambda}$ associated with `dark energy.' In fact, total wavefunction symmetrization for BE ground states of either photons or gluons may be expected to result in effective attractive potentials.\cite{Pathria} 

Let us then proceed to estimate the critical temperature, $T_c$, of BE condensation of massive bosons. Assuming for their ordinary phase a relativistic energy-momentum free-particle relation $\epsilon \simeq c p$, the density for just about all their excited states at $z=1$ fugacity is\cite{Pathria}
\begin{equation}\label{excitedstates}
n_{ex} \simeq g_s \frac{4 \pi (k_B T)^3} {(h c)^3} 2 \zeta(3) .
\end{equation}  
Here $g_s = 3$ is the $S =1$ spin degeneracy for massive bosons and $\zeta(3) \simeq 1.202$. Equating $n_{ex}$ to $n_g$ of \Eq{numberdensity} determines the threshold at which bosons begin to condensate `en masse' into the ground state. That yields

\begin{equation}\label{criticalenergy}
k_B T_c \sim \frac{h c} {d_g} {[8 \pi g_s \zeta(3)]}^{-1/3} \simeq 17.25 MeV  
\end{equation}  
and a critical temperature
\begin{equation}\label{criticaltemperature}
 T_c \sim 2 \mathrm{x} 10^{11}K .
\end{equation}   

This estimate of massive boson BE condensation reaches to the end of the quark epoch, at about $10^{-6}$ s and $10^{12}$ K, when quarks and gluons began to confine and form hadrons. Once again, that may provide a tantalizing clue. This massive gluon BE condensate then behaves like that of a near-ideal Bose gas, which differs from its antecedent, a quark-gluon plasma, which is thought to behave like a near-ideal Fermi liquid. Amazingly, both energy regions are already accessible to current accelerators, such as LHC. 

Actually, the most problematic aspect of this BE condensate scenario is the presumption that massive gluons would be nearly not interacting among themselves. That may be so in the asymptotic freedom limit. Yet, gluons carry color, except for color singlets. However, gluons in a color-singlet state are generally excluded, as they would mediate long-range strong interactions among hadrons, which are not observed.\cite{Quigg, Mandl} Thus, other kinds of gluon complexes may have to be considered for the BE ground state condensate that I propose, or that model should be abandoned. On the other hand, there would be no such problem for a BE condensate of massive photons, which carry no electric charge and do not interact among themselves. 

Also puzzling may be the idea that massive boson BE condensates may contribute to `dark matter,' $\rho_{dm}$, in addition to `dark energy,' $\rho_{\Lambda}$. However, $\rho_{dm}$ and $\rho_{\Lambda}$ have completely different distributions and dependence on the scale function $R(t)$ of the FLRW geometry of the observable universe.\cite{Hobson, Schutz2Ed} Introducing $\rho_{dm}$ into my original scenario may require major revisions, which may still be possible, however.

In conclusion, I have proposed a conjecture of non-zero photon and gluon bare masses based on a cosmological scale. That entails the formation of massive boson Bose-Einstein condensates at rest in a comoving Friedmann-Lemaitre-Robertson-Walker geometry. I have derived tantalizing results that set the inter-particle distance in these condensates almost at the nucleon scale and a corresponding critical temperature of condensation almost at the end of the quark epoch or the beginning of quark-gluon confinement. Gluon complexes may have to be considered for their Bose-Einstein ground state, whereas a condensate of single massive free photons may more readily form. There are of course much more sophisticated theories of matters that I address, but those are further beyond the standard model of elementary particles and the scope of this paper. In particular, ultralight bosons that are considered as candidates for `dark matter' are of a quite different nature.\cite{Brito} 



\bibliographystyle{apsrev4-1}

\begin{thebibliography}{46}%
\makeatletter
\providecommand \@ifxundefined [1]{%
 \@ifx{#1\undefined}
}%
\providecommand \@ifnum [1]{%
 \ifnum #1\expandafter \@firstoftwo
 \else \expandafter \@secondoftwo
 \fi
}%
\providecommand \@ifx [1]{%
 \ifx #1\expandafter \@firstoftwo
 \else \expandafter \@secondoftwo
 \fi
}%
\providecommand \natexlab [1]{#1}%
\providecommand \enquote  [1]{``#1''}%
\providecommand \bibnamefont  [1]{#1}%
\providecommand \bibfnamefont [1]{#1}%
\providecommand \citenamefont [1]{#1}%
\providecommand \href@noop [0]{\@secondoftwo}%
\providecommand \href [0]{\begingroup \@sanitize@url \@href}%
\providecommand \@href[1]{\@@startlink{#1}\@@href}%
\providecommand \@@href[1]{\endgroup#1\@@endlink}%
\providecommand \@sanitize@url [0]{\catcode `\\12\catcode `\$12\catcode
  `\&12\catcode `\#12\catcode `\^12\catcode `\_12\catcode `\%12\relax}%
\providecommand \@@startlink[1]{}%
\providecommand \@@endlink[0]{}%
\providecommand \url  [0]{\begingroup\@sanitize@url \@url }%
\providecommand \@url [1]{\endgroup\@href {#1}{\urlprefix }}%
\providecommand \urlprefix  [0]{URL }%
\providecommand \Eprint [0]{\href }%
\providecommand \doibase [0]{http://dx.doi.org/}%
\providecommand \selectlanguage [0]{\@gobble}%
\providecommand \bibinfo  [0]{\@secondoftwo}%
\providecommand \bibfield  [0]{\@secondoftwo}%
\providecommand \translation [1]{[#1]}%
\providecommand \BibitemOpen [0]{}%
\providecommand \bibitemStop [0]{}%
\providecommand \bibitemNoStop [0]{.\EOS\space}%
\providecommand \EOS [0]{\spacefactor3000\relax}%
\providecommand \BibitemShut  [1]{\csname bibitem#1\endcsname}%
\let\auto@bib@innerbib\@empty

\bibitem[Hobson(2006)]{Hobson}
Hobson, M. P., Efstathiou, G. P., Lasenby, A. N., General Relativity: An Introduction for Physicists, ~Cambridge University Press, 2006.%

\bibitem[Schutz(2009)]{Schutz2Ed}
Schutz, B. F., A First Course in General Relativity, 2nd Ed., ~Cambridge University Press, 2009.%

\bibitem[Rindler(1979)]{Rindler} 
Rindler, W., Essential Relativity: Special, General, and Cosmological, ~Revised 2nd Ed., ~Springer-Verlag, 1979.%

\bibitem[Quigg(2013)]{Quigg} 
Quigg, C., Gauge Theories of the Strong, Weak, and Electromagnetic Interactions, 2nd Ed., Princeton University Press, NJ, 2013.%

\bibitem[Mandl(2010)]{Mandl} 
Mandl, F., Shaw, G., Quantum Field Theory, 2nd Ed., Wiley, NY, 2010.%

\bibitem[Jackson(2013)]{Jackson} 
Jackson, J. D., Classical Electrodynamics, 3rd Ed., Wiley, NY, 1999.%

\bibitem[Dolgov(1981)]{Dolgov}
Dolgov, A. D., Zeldovich, Ya. B., Cosmology and elementary particles, Rev. Mod. Phys. \textbf{53} (1), 1-39, 1981; https://doi.org/10.1103/RevModPhys.53.1%

\bibitem[Pani(2012)]{Pani}
Pani, P., \textit{et al.}, Black-Hole bombs and photon-mass Bounds, Phys. Rev. Lett. \textbf{109}, 131102, 2012; https://link.aps.org/doi/10.1103/PhysRevLett.109.131102.%

\bibitem[Pathria(2011)]{Pathria}
Pathria, R. K., Beale, P. D., Statistical Mechanics, 3rd Ed., Academic Press, Elsevier, Amsterdam, 2011.%

\bibitem[Brito(2017)]{Brito}
Brito, R., \textit{et al.}, Gravitational wave searches for ultralight bosons with LIGO and LISA, Phys. Rev. D. \textbf{96}, 064050 (2017); DOI: 10.1103/PhysRevD.96.064050; arXiv:1706.06311 [gr-qc].%














\end{thebibliography}


\end{document}